\documentclass[12pt]{iopart}
\usepackage{graphicx}

\begin{document}
\bibliographystyle{unsrt}

\paper[]{Design of quasi-lateral $p$-$n$ junction for optical spin-detection in low-dimensional systems}

\author{Bernd K\"astner\dag\ \footnote[3]{To whom correspondence should be addressed (bk221@cam.ac.uk)}, D. G. Hasko\dag, and D. A. Williams\ddag}

\address{\dag\ Microelectronics Research Centre, Cavendish Laboratory, University of Cambridge, Madingley Road, Cambridge CB3 0HE, UK.}

\address{\ddag\ Hitachi Cambridge Laboratory, Madingley Road, Cambridge CB3 0HE, UK.}

\begin{abstract}

A technology is reviewed which allows one to produce quasi-lateral 2D electron and hole gas junctions of arbitrary shape. It may be implemented in a variety of semiconductor heterostructures. Here we concentrate on its realization in the GaAs/AlGaAs material system and discuss the possibility to use this structure for optical spin detection in low-dimensional systems.

\end{abstract}




\section{Introduction}

The study of electron spin in materials, in order to better
understand its behavior, offers the hope of developing an entirely
new generation of microelectronic devices. However, direct 
detection of the electron spin is rather difficult, since
the corresponding magnetic moment is too small. Therefore, spin
detection can only be achieved by coupling it to more accessible
parameters. One possibility is to apply \emph{optical} methods
for the spin detection. Here, the ability to couple the electron
spin to optical photons is exploited \cite{aronov1PAIII} by
studying the degree of circular polarization of the
electroluminescence emitted by a spin sensitive light emitting
diode (LED). The optical detection of spin is much more accessible
than other methods. Several groups have used this technique to
determine the degree of polarization of the carriers injected into
a non-magnetic semiconductor \cite{fiederling1PBII, ohno1PBII,
zhu1PAIII}.

However, the combination of the concept of a light 
emitting diode (LED) and the possibility of
low-dimensional transport is relatively complex. It requires the use of
planar-geometry schemes, which have recently been developed
\cite{vaccaro1PBII, kaestner1PBII, cecchini1PBII}. Only a few
implementations exist of $p$-$n$ junctions with lateral geometries
and were, until recently, based on crystal plane dependent doping
\cite{miller2PBII}. Using this technique, lateral low-dimensional
$p$-$n$ junctions were demonstrated by T. Yamamoto
\cite{yamamoto1PBII}. However, the fabrication of such devices
require special growing facilities and is relatively complex.
Also, the radiative recombination efficiency might be very low
since the active layer is Si-delta doped. Here we review an 
alternative way to produce a lateral junction between a
\emph{two-dimensional electron gas} (2DEG) and \emph{hole gas}
(2DHG) which has recently been realized.

The following sections will first explain the basic idea of the
structure, and its limits. A one-dimensional Poisson-solver will
be used next to predict the correct structure parameters relevant
at low temperatures, where the device is intended to be used. Then
a two-dimensional simulation will be presented, showing the
potential distribution at $T=300\,$K in the cross section. In this
way, some information of the lateral dimensions of the junction is
provided, which is assumed to be of the same order as for low
temperatures.

\section{Basic Idea}

The way in which confinement was achieved is based on the method
of modulation doping, where two materials with almost identical
lattice constants but different band gaps are grown on top of each
other to form a heterojunction. Here the material
GaAs/Al$_x$Ga$_{1-x}$As will be employed. If the material with the
larger bandgap (AlGaAs) is doped with shallow donors or acceptors,
a phenomenon called \emph{band bending} occurs. Due to band
bending electrons or holes are confined by an approximately
triangular potential barrier near the interface and form a 2DEG or
2DHG. The modulation doping will be explained in the next section.

The structure will be designed so that radiative recombination
takes place in GaAs. According to the selection rules
\cite{bastard1PBII} in zinc-blende structures, such as GaAs,
excited polarized electrons lead to circularly polarized
electroluminescence. It therefore satisfies the condition to be
used for spin-detection.

The task is now to produce a junction between the two-dimensional
(2D) electrons and holes such that under forward bias
recombination takes place in a well defined GaAs region. In the
case of electrons, the AlGaAs would have to be doped with donors
and for holes with acceptors, resulting in carrier transport
parallel to the AlGaAs/GaAs interface. This means that a junction
can only be formed in a lateral fashion.

When material such as GaAs is grown, for instance by molecular
beam epitaxy, only custom designed facilities allow a lateral
variation of material to be deposited. Therefore a different approach is used here. 
The low bandgap material is
sandwiched between two high bandgap materials, one acceptor doped
and one donor doped. In the as-grown state, the potential electron
gas near one of the two interfaces is fully depleted, and only the
2D hole gas is generated near the other interface. By removing the
acceptor doped high bandgap material the 2D electron gas can now
develop at the interface to the remaining high bandgap material. A
junction can thus be generated by removing the high bandgap
material only over parts of the wafer.

The narrower one can make the sandwiched low bandgap material, the
closer would be the vertical off-set of the electron and hole gas
in a possible junction. However, since, as explained later in this
chapter, band bending is also caused by the surface exposed to air
after removing one of the two high bandgap regions, there is a
limit as to how thin one can make the low bandgap region.
Therefore, a careful tuning of various structure parameters is
needed and will be discussed below.

\begin{figure}[tb]
\begin{center}
\includegraphics[width=10cm, bb=-88 371 683 590]{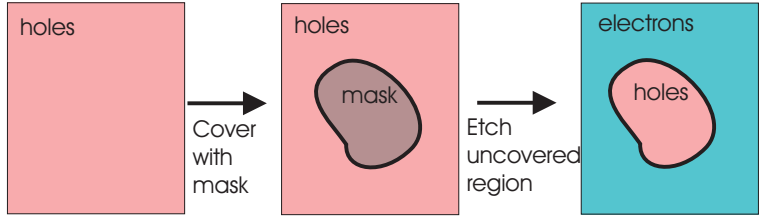} \hspace{.5in}
\end{center}
\caption{\label{figEtchSchem} Schematic view on the surface of an
arbitrary wafer, which initially contains a sheet of holes
parallel to the surface (indicated as red). By masking certain
areas and subsequent etching the sheet of holes disappears and
instead a sheet of electrons is formed parallel to the surface.}
\end{figure}

The advantage of this design is that by applying etch masks any
shape of electron- and hole-sheets can be defined on the same
wafer, as illustrated in Figure \ref{figEtchSchem}.

\section{Modulation Doping}

The main task in the design of the structure is to apply the
modulation doping method by tuning parameters, such as bandgap,
doping and dimensions. This section reviews modulation doping
based on the book by Bastard \cite{bastard1PBII}.

\begin{figure}[tb]
\begin{center}
\includegraphics[width=10cm, bb=82 602 396 782]{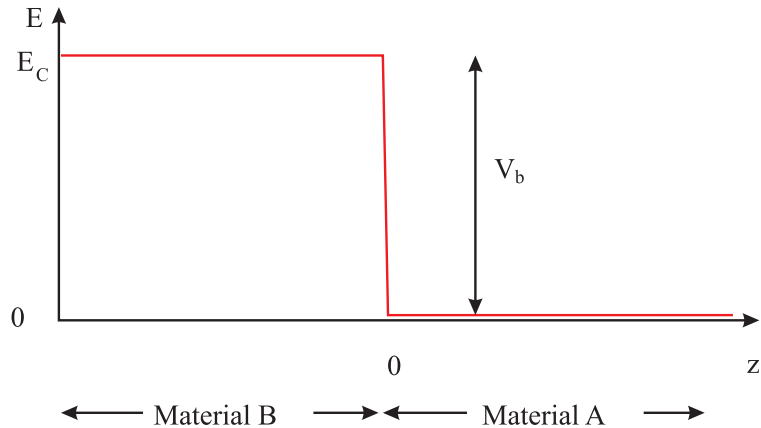}
\end{center}
\caption{\label{figFlatEcProfile} Conduction band profile of an
abrupt heterojunction under flat band conditions.}
\end{figure}

Consider an abrupt heterojunction between the two materials A and
B, having different band gaps. A difference in two material
bandgaps creates conduction and valence band discontinuities.
Under flat-band conditions (unperturbed case) a possible
conduction band profile is shown in Figure \ref{figFlatEcProfile}.
Suppose material B contains impurities (for simplicity $n$-type)
and material A is intrinsic. One may regard an electron in the
presence of a donor impurity of charge $+e$ within the medium of
the semiconductor, as a particle of charge $-e$ and mass $m^\ast$.
This is precisely the problem of a hydrogen atom, except that the
product $-e^2$ of the nuclear and electronic charges must be
replaced by -$e^2/\epsilon$, and the free electron mass $m$, by
$m^\ast$ (\cite{ashcroft1PBII}, p.577 onwards). In almost all
cases the binding energy $R^\ast$ of an electron to a donor
impurity is small compared with the energy gap $E_g$ of the
semiconductor.

However, the electrons are bound relative to the conduction band
edge of the B material. Therefore, their energies are $\approx
V_b-R^\ast$ above the onset of the heterojunction continuum. At $T
= 0 K$ the electrons would all be frozen on the impurity site but
this situation is unstable since it does not ensure the equality
of the Fermi level $E_F$ in both sides of the heterojunction. In
material B it should lie between the donor levels and the
conduction band edge, whereas in material A it should be negative
since there are no free carriers.

To relax to thermal equilibrium, some of the carriers, assumed to
be trapped at time $t=0$ onto the donor sites, tunnel or are
emitted thermionically in the A material. By emitting phonons
those electrons relax to $E = 0$ (in a time scale of $\approx
10^{-12}s)$. The reverse process is unlikely, and in fact at $T =
0 K$ impossible since no phonons are available to match the energy
difference $V_b - R^\ast$ and tunnelling would correspond to the
capture of an electron moving quickly in the layer plane, which is
both inefficient and unlikely. Therefore, doping only the barrier
acting material B, a spontaneous and irreversible charge transfer
to the well acting material A is induced.

\begin{figure}[tb]
\begin{center}
\includegraphics[width=10cm, bb=92 567 426 788]{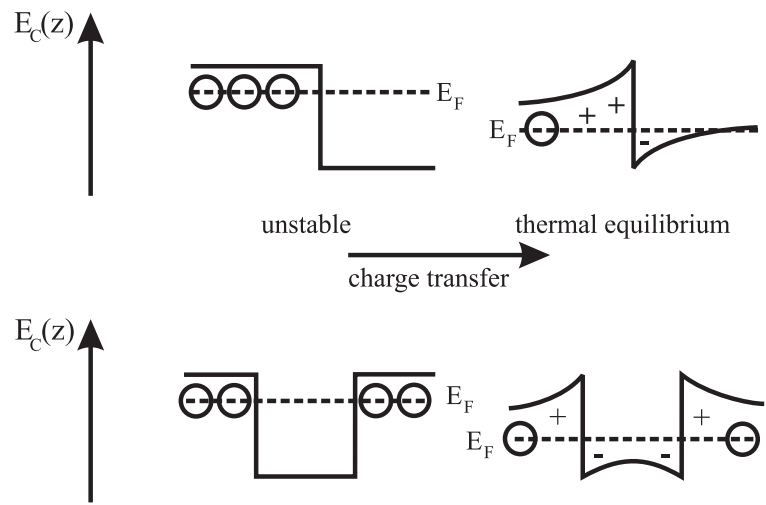} \hspace{.5in}
\end{center}
\caption{\label{figModDop} Charge transfer in modulation doped
single and double heterostructures. Circles denote the approximate
energy level of  neutral donors and the plus and minus sign stand
for ionized donors and electrons, respectively.}
\end{figure}

The spatial separation of electrons and their parent donors leads
to band bending due to dipole formation between positively ionized
donors and the electron on the other side of the heterojunction
interface. This band bending only depends on $z$, if one averages
the donor distribution in the layer plane. This results in a
quasi-triangular potential near the interface as shown in Figure
\ref{figModDop}, leading to bound states $E_1, E_2, ...$ for the
$z$-motion. If the energy spacings are much larger than the
thermal or collisional broadenings, the carrier motion becomes
effectively quasi two-dimensional. One can repeat the reasoning
with acceptor-doped barrier material: In this case the free
carriers in the well-acting material are holes, which are bound
and may form a quasi-two dimensional gas.

\section{Asymmetric Modulation Doping}

It is possible to replace Ga atoms in GaAs with Al to make the
ternary material $\mathrm{Ga_{1-x}Al_x As}$ without any
significant change in the electronic arrangement of the crystal
lattice. At the same time the band gap changes from 1.43$\,$eV
(GaAs) up to 2.15$\,$eV (temperature dependent). This material
system is therefore suitable for implementing the modulation
doping technique.

\begin{figure}[tb]
\begin{center}
\includegraphics[width=10cm]{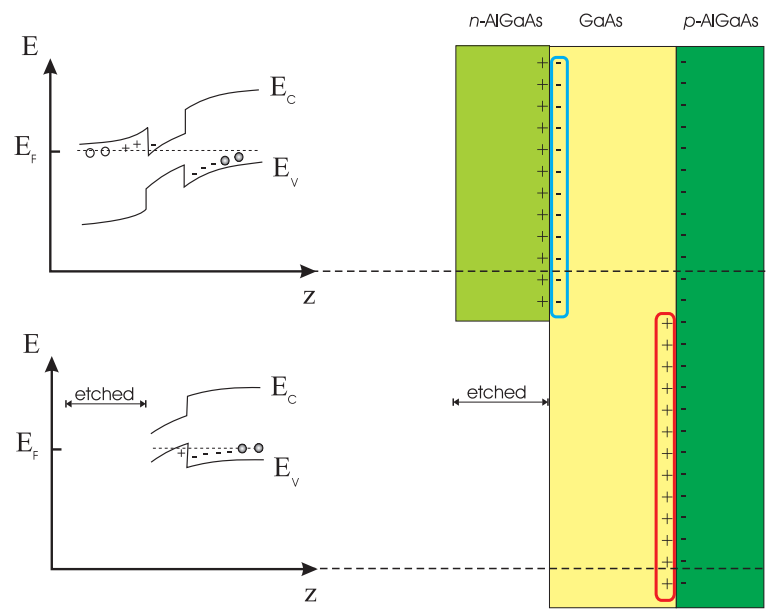} \hspace{.5in}
\end{center}
\caption{\label{figAsModDop} Asymmetrically modulation doped
double heterostructures. In the left part, open and filled circles
denote the approximate energy level of  neutral donors and
acceptors, respectively. The plus and minus sign stand for ionized
donors, and holes on the one hand, and ionized acceptors and
electrons on the other hand, respectively. On the right, the
schematic cross-section of the partly etched structure is shown.
The dashed lines indicate the position which the band diagrams on
the left part correspond to. Inside the GaAs region there are free
electrons (blue) and holes (red).}
\end{figure}

\clearpage

As explained above, in order to produce a lateral junction between
electrons and holes, the low-bandgap material, which will be GaAs,
is sandwiched between high bandgap material (Al$_x$Ga$_{1-x}$As)
that on one side is acceptor and the other is donor doped. The
situation is shown in the left part of Figure \ref{figAsModDop}.
For the diagram on top there are three possible cases: first,
both, electrons and hole populate GaAs, secondly, only one type is
present and thirdly, none. The desired case is the second one,
where only one type of carrier is present, with the additional
constraint that if the barrier-acting AlGaAs on one side is
removed the other carrier type populates the GaAs material, as
seen in the bottom diagram of Figure \ref{figAsModDop}.

In this way, a junction between a two-dimensional electron gas,
and a two-dimensional hole gas is produced at the edge between the
regions where the top barrier-acting material has been removed.
This is shown schematically in the right part of Figure
\ref{figAsModDop}.

\section{Fermi-Level Pinning at the Surface}
\label{secFermi}

As indicated in the left part of Figure \ref{figAsModDop} the
Fermi level $E_F$ at the left surface lies in the middle of the
band gap. This pinning of the Fermi level at a fixed value (in
this case at mid-gap) is caused by surface states. The closer the
surface the heavier one has to dope the barrier-acting material in
order to compensate for these surface effects. Therefore,
Fermi-level pinning has to be taken into account when designing a
structure like this.

The basic effect of cleaving and removing half of the crystal is
to break bonds, which are left as singly occupied atomic orbitals
known as \emph{dangling bonds}. A monolayer of such bonds should
form a two-dimensional energy band $E_s(\mathbf{k}_\parallel)$
because of the translational symmetry along the surface. Those
surface bands may lie, at least in part, in the forbidden energy
gap. In this case one talks about \emph{surface states}.

States in the forbidden gap will cause the Fermi level to change
as one approaches the surface, in a way explained below. Clean,
cleaved GaAs is free of surface states such that the position of
the bulk bands with respect to the Fermi level is constant from
inside the crystal all the way to the surface (p.488
\cite{cardona1PBII}). This suggests that surface states do not
necessarily arise for mechanical reasons. However, there is a
chemical origin for GaAs surface states. This is revealed when
cleaved GaAs is exposed to Oxygen. Photoelectron spectroscopy
studies show that the surface Fermi level of GaAs (110) dosed with
O$_2$ is pinned for no more than 5\% of oxygen surface coverage
\cite{spicer1PM}. This revealed the appearance of elemental As at
the surface \cite{thurmond1PM, spicer1PM, ping1PM, yi1PM},
suggesting that the GaAs was being chemically dissociated by the
oxygen. This resulted in new electronic states (acceptors and
donors) close to the exposed GaAs surface and within the bandgap.

The surface state density is generally quite high ($10^{12} -
10^{13}$/eV cm$^2$) occurring at 0.4 to 0.5 eV above the valence
band energy \cite{hasegawa1PM}. An $n$-type doping producing
$\approx 10^{18}$ electrons/cm$^3$ leads to a carrier density at
the surface of $\approx 10^{10}$ electrons/cm$^2$. Hence, the
Fermi level can only penetrate into the surface bands by an amount
\[
 \frac{10^{10} \mathrm{/cm}^2}{10^{12} \mathrm{/eV cm}^2}
   = 10^{-2} \mathrm{eV}.
\]
This means that the Fermi level barely penetrates into the band of
empty surface states. This is the origin of the phenomenon known
as Fermi level \emph{pinning}. It leads to the band bending
displayed in the right diagram of Figure \ref{figAsModDop}. This
surface band bending results from the fact that in equilibrium the
Fermi level $E_F$ must remain constant from bulk to surface: the
band edges must then vary so as to be compatible with $E_F$
pinning at the surface and the bulk well away from it.

\section{Wafer Design Based on a 1D Poisson Simulation}
\label{sec1DPoi}

To predict which acceptor and donor densities and which channel
width would result in a conducting lateral junction after the etch
process the band-diagram as well as the charge density was
simulated using a 1D Poisson solver \cite{tan1POI}. It also takes
quantized states into account, using the one-dimensional
Schr\"odinger equation:
\begin{equation}
 -\frac{\hbar^2}{2}\frac{d}{dz}\left( \frac{1}{m^\ast} \frac{d}{dz}
 \right)\psi(z) + U(z)\psi(z) = E \psi(z),
 \label{eqn1DSG}
\end{equation}
where $\psi$ is the wave function, $E$ is the total energy, $U$
the potential energy, $\hbar$ is Planck's constant divided by
$2\pi$, and $m^\ast$ is the effective mass. The potential energy
$U$ may be set to be equal to the conduction band energy $E_C$.

For the AlGaAs system the conduction band and valence band
effective masses, for electrons, light holes and heavy holes are
given, respectively, in terms of the free electron mass $m_e =
9.10956 \times 10^{-31}\,$kg by:
\begin{eqnarray}
 m_e^* &=&0.067\,m_e, \nonumber \\
 m_{lh}^* &=& 0.082\,m_e, \nonumber \\
 m_{hh}^* &=& 0.45\,m_e.
 \label{eqnEffMass}
\end{eqnarray}

The wavefunction $\psi$ in (\ref{eqn1DSG}) and the electron
density $n$ are related by
\begin{equation}
 n(z)=\sum^l_{k=1} \psi^\ast_k(z) \psi_k(z) n_k,
 \label{eqnElDen}
\end{equation}
where $l$ is the number of bound states, and $n_k$ is the electron
occupation for each state. In general, the number of carriers in
thermal equilibrium can be calculated from the density of levels
$g_c(E)$ in the conduction band and $g_v(E)$ in the valence band
at temperature $T$ by
\begin{eqnarray}
  \label{eqnCarrDen}
  n_c(T) & = & \int_{E_c}^\infty dE g_c(E)\frac{1}{e^{(E-E_F)/k_B T} +
  1},\nonumber\\
  p_v(T) & = & \int^{E_v}_{-\infty} dE g_v(E)\left(1-\frac{1}{e^{(E-E_F)/k_B T} +
  1}\right)\nonumber\\
    & = & \int^{E_v}_{-\infty} dE g_v(E)\frac{1}{e^{(E_F-E)/k_B T} +
  1}.
\end{eqnarray}
In the case of 2D quantized levels each sublevel has a constant
density of states $g_{c,v}=m^*_{e,h}/\pi \hbar^2$. The electron
occupation for each sublevel can therefore be expressed by
\begin{equation}
 n_k = \frac{m_e^\ast}{\pi \hbar^2} \int^\infty_{E_k} \frac{1}{1+e^{(E-E_F)/k
 T}}dE,
 \label{eqnElCon}
\end{equation}
where $E_k$ is the eigenenergy. The hole density $p(z)$ can be
calculated in a similar way, by setting the potential energy $U$
equal to the negative valence band energy $E_V$.

In a quantum well of arbitrary potential energy profile, the
spatial variation of $E_C$ is related to the electrostatic
potential $\phi$ as follows:
\begin{equation}
 E_C(z) = - q \phi (z) + \Delta E_C (z),
 \label{eqnPotEnRel}
\end{equation}
where $\phi$ is the electrostatic potential and $\Delta E_C$ is
the pseudopotential energy due to the band offset at the
heterointerface. How the band gap difference $E_{g1}-E_{g2}$ of
the regions 1 and 2 is distributed between the conduction and
valence bands has a large impact on the charge transport in these
heterodevices. The conduction band discontinuity is specified:
$\Delta E_C = (E_{g2}-E_{g1})*0.6$. Knowing the band gap energies,
the spatial variation of $E_V$ can also be derived. There are
three primary conduction bands in the AlGaAs system that depending
on the Al mole fraction $x$, determine the bandgap. These are
named $\Gamma$, L and X. The default bandgaps for each of these
conduction band valleys are as follows:
\begin{eqnarray}
 E_{g\Gamma} & = & E_{g0} + x (0.574+0.055 x) \nonumber\\
 E_{gL}      & = & 1.734 + x (0.574 + 0.055 x)\nonumber \\
 E_{gX}      & = & 1.911 + x(0.005 + 0.245 x). \nonumber
\end{eqnarray}
$E_{g0}$ is the bandgap at $300\,$K and set $1.422\,$eV. The
temperature dependence of the bandgap is calculated according to
\begin{equation}
  E_g(T) = E_g(300\,\mathrm{K}) + \alpha \left[
           \frac{(300\,\mathrm{K})^2}{300\,\mathrm{K}+\beta}
           -\frac{T^2}{T + \beta}\right],
\end{equation}
where the constants are set as $\alpha=5.405\times10^{-4}\,$eV/K
and $\beta = 204\,$K. $E_g(300\,\mathrm{K})$ is taken as the
minimum of $E_{g\Gamma}$, $E_{gL}$, and $E_{gX}$.

The electrostatic potential $\phi(z)$ can be derived from the
charge carrier densities $n$ and $p$ via the one-dimensional
Poisson equation
\begin{equation}
 \frac{d}{dz}\left( \epsilon_s (z) \frac{d}{dz}\right) \phi (z) =
 \frac{- q [N_D^+(z) - N_A^- (z) - n(z) + p(z)]}{\epsilon_0},
 \label{eqnEStatPot}
\end{equation}
where $\epsilon_s$ is the dielectric constant, $N_D^+$ is the
ionized donor concentration, and $N_A^-$ the ionized acceptor
concentration. The static dielectric constant $\epsilon_s$ is
$z$-dependent, since it varies with the Al-mole fraction $x$. For
the AlGaAs-system it is given by
\[
 \epsilon_{\mathrm Al_{x}Ga_{1-x}As} = 13.8 + 2.9 x.
\]

An iteration procedure is used to obtain self-consistent solutions
for (\ref{eqn1DSG}) and (\ref{eqnEStatPot}). Starting with a trial
potential energy $U(z)$, the wave function, and their
corresponding eigenenergies, $E_k$ can be used to calculate the
carrier density distribution $n(z)$ and $p(z)$ using equations
(\ref{eqnElDen}) and (\ref{eqnElCon}). Equation
(\ref{eqnEStatPot}) can then be used to calculate $\phi (z)$. The
new potential energy $U(z)$ is then obtained from
(\ref{eqnPotEnRel}). The subsequent iteration will yield the final
self-consistent solutions for $U(z)$, $n(z)$, and $p(z)$, which
satisfy certain error criteria.

\begin{table}[t]
\begin{center}
 \begin{tabular}{llp{2cm}ll}\hline
  No. & Material & Thickness in nm & Doping in cm$^{-3}$ &
  remark\\\hline\hline
  1 & GaAs & 5 & $N_a = 1.0\times 10^{19}$ & capping layer\\
  2 & Al$_{0.5}$Ga$_{0.5}$As & 10 & $N_a = 1.0\times 10^{18}$ & \\
  3 & Al$_{0.5}$Ga$_{0.5}$As & 10 & $N_a = 6.0\times 10^{18}$ & \\
  4 & Al$_{0.5}$Ga$_{0.5}$As & 2 & undoped & spacer layer\\
  5 & GaAs & 30 & undoped & conductive channel\\
  6 & Al$_{0.4}$Ga$_{0.6}$As & 3 & undoped & spacer layer\\
  7 & Al$_{0.4}$Ga$_{0.6}$As & 5 & $N_d = 3.9 \times 10^{18}$ & \\
  8 & Al$_{0.4}$Ga$_{0.6}$As & 20 & undoped & \\
  9 & Al$_{0.3}$Ga$_{0.7}$As & 300 & undoped & buffer layer \\
  10 & AlGaAs/GaAs & 500 & undoped & Superlattice \\
  11 & GaAs & - & undoped & Substrate\\\hline
 \end{tabular}
 \caption{\label{tabAIdeal}Wafer configuration of a possible
 implementation of lateral junction by etching layers 1 to 4.}
\end{center}
\end{table}

This algorithm is now used to simulate the implementation of the
lateral junction into the AlGaAs/GaAs material system. The
temperature was assumed to be 4.2$\,$K. A possible configuration
is shown in Table \ref{tabAIdeal}. Several fabrication issues have
been taken into account in the structure design already. Layer 1
is needed in order to protect the Al$_{0.5}$Ga$_{0.5}$As layer
underneath from oxidizing. In addition, it was heavily doped which
facilitates ohmic contact formation. The acceptor doping was
chosen to be above the donor layers since the highly diffusive
Beryllium was used as the acceptor dopant. Furthermore, the
acceptor layers will have to be removed with high selectivity over
the undoped GaAs layer (layer number 5). A selective etch exists
which removes Al$_{x}$Ga$_{1-x}$As for $x>0.4$ over $x<0.4$ \cite{wu1POI}. 
In order to achieve a high
selectivity, the Aluminium content was chosen to be 50\%.

The advantage of modulation doping, as far as impurity scattering
is concerned, is the improvement in mobility due to the spatial
separation between the carriers and their parent donors/acceptors.
This spatial separation can be further enhanced by inserting a
spacer layer, which is a nominally undoped part of the barrier,
between the donor/acceptor and the 2DEG/2DHG. Therefore layers 4
and 6 have been inserted to act as spacer layers.

\begin{figure}[tb]
\begin{center}
\includegraphics[width=12cm]{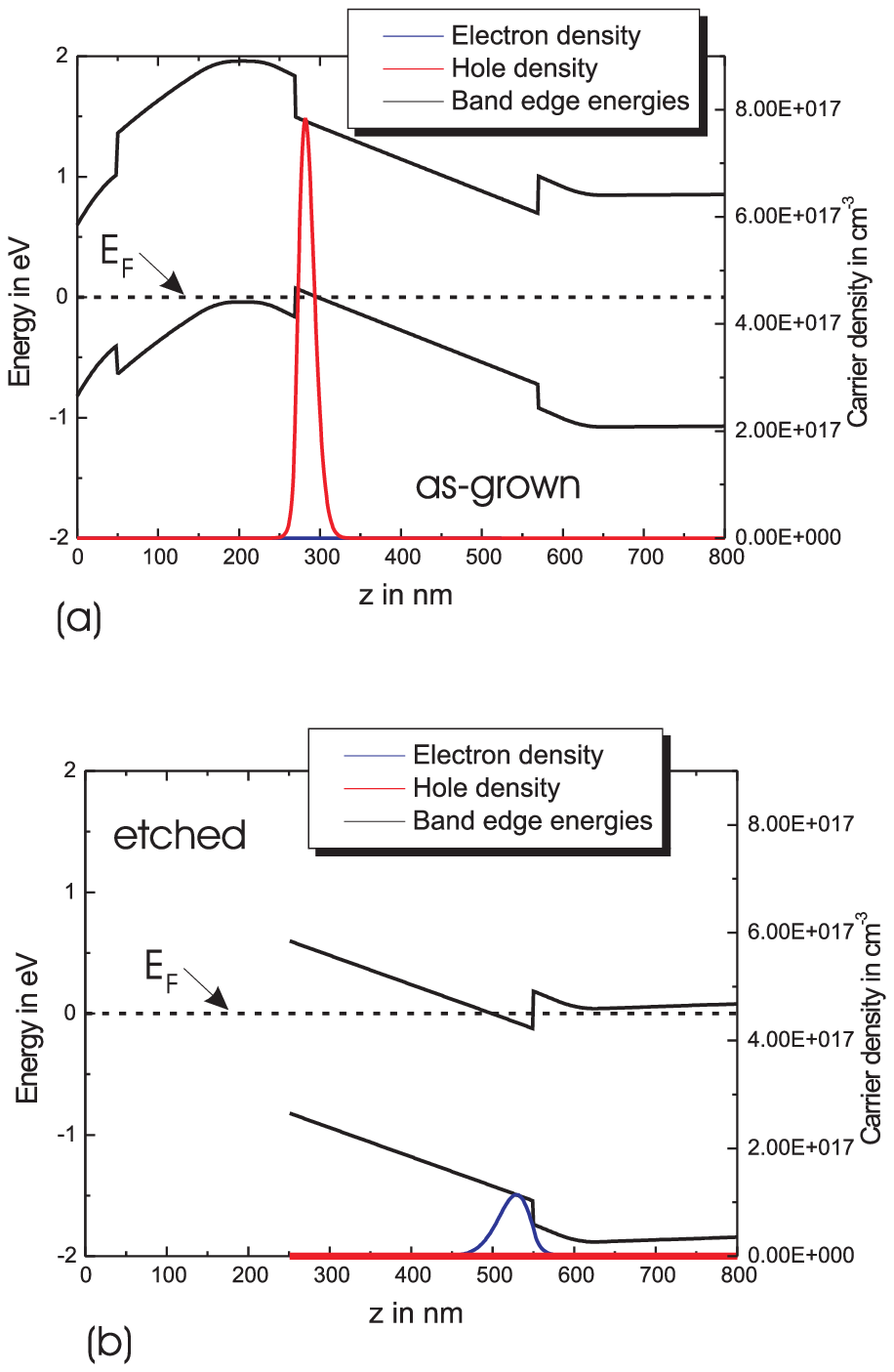}
\vspace{-0.7cm}
\end{center}
\caption{\label{figAIdeal} One-dimensional Poisson simulation of
an asymmetrically modulation doped wafer, changing the carrier
type when etched. The as-grown state is shown in part ($a$), where
the carriers are $p$-type. When the top layers are removed, the
carriers change to $n$-type.}
\end{figure}

\clearpage

Figure \ref{figAIdeal} shows the simulated band diagram at
4.2$\,$K of the etched and the as-grown wafer. The potential
energy is normalized so that $E_F = 0\,$eV. Etching away layers 1
to 4 causes a significant change. In the as-grown state, $E_F$
lies well below the conduction band edge everywhere on the
$z$-axis. At the interface between layer 4 and 5 a sheet of holes
is generated in the GaAs channel, as indicated by the red curve in
Figure \ref{figAIdeal}. This situation changes if the top 4 layers
are removed. In this case $E_F$ lies above the valence band edge
everywhere on the $z$-axis. All free holes are depleted, but at
the interface between layers 5 and 6 there is a formation of free
electrons.

Even though the simulation of the structure does show the correct
behavior, in reality varying activation levels of dopants in different growing chambers 
would require to start with a design which would allow
for deviations from the simulation. A structure subject to optimization
should at least be conductive and show carrier type alteration
when etched. Other issues, such as parallel conduction and the
actual vertical distance between the sheet of electrons and holes,
should be considered after the device operation was demonstrated.
The parameters of two possible wafer structures are
shown in Table \ref{tabAReal} and have been tested in \cite{kaestner2PBII, kaestner3, kaestner}.

\begin{table}
\begin{center}
 \begin{tabular}{|l|l|l|r|l|}\hline
  No. & \multicolumn{2}{c|}{Material} & Thickness & Doping in
  cm$^{-3}$\\\cline{2-3}
  & Wafer 2 & Wafer 1 & in nm & \\\hline\hline
  1 & GaAs & GaAs & 5 & $N_a = 1.0\times 10^{19}$ \\
  2 & Al$_{0.5}$Ga$_{0.5}$As & Al$_{0.5}$Ga$_{0.5}$As & 50 & $N_a = 8.0\times 10^{18}$ \\
  3 & Al$_{0.3}$Ga$_{0.7}$As & Al$_{0.5}$Ga$_{0.5}$As & 3 & undoped \\
  4 & GaAs & GaAs & 90 & undoped \\
  5 & Al$_{0.3}$Ga$_{0.7}$As & Al$_{0.3}$Ga$_{0.7}$As & 5 & undoped
  \\\cline{4-5}
  6 & Al$_{0.3}$Ga$_{0.7}$As & Al$_{0.3}$Ga$_{0.7}$As
    & \multicolumn{2}{l|}{$\delta$-doped  $N_d = 5.0 \times 10^{12}$ cm$^{-2}$}
    \\\cline{4-5}
  7 & Al$_{0.3}$Ga$_{0.7}$As & Al$_{0.3}$Ga$_{0.7}$As & 300 & undoped  \\
  8 & AlGaAs/GaAs SL & AlGaAs/GaAs SL & 500 & undoped \\
  9 & GaAs & GaAs & - & undoped \\\hline
 \end{tabular}
 \caption{\label{tabAReal}Parameter of wafers 1 and 2.}
\end{center}
\end{table}

Wafer 1 and 2 only differ in layer 3. This difference has the
following two consequences. Firstly, the etch will remove
Al$_{0.5}$Ga$_{0.5}$As, but not the thin Al$_{0.3}$Ga$_{0.7}$As
layer, which was intended to passivate the surface and to cause
un-pinning of the Fermi-level \cite{parikh1POI} (not observed).
Secondly, the confinement at the interface between layer 2 and 3
will be stronger in the case of wafer 1, since the bandgap offset
will be larger in this case.

\begin{figure}[tb]
\begin{center}
\includegraphics[width=12cm]{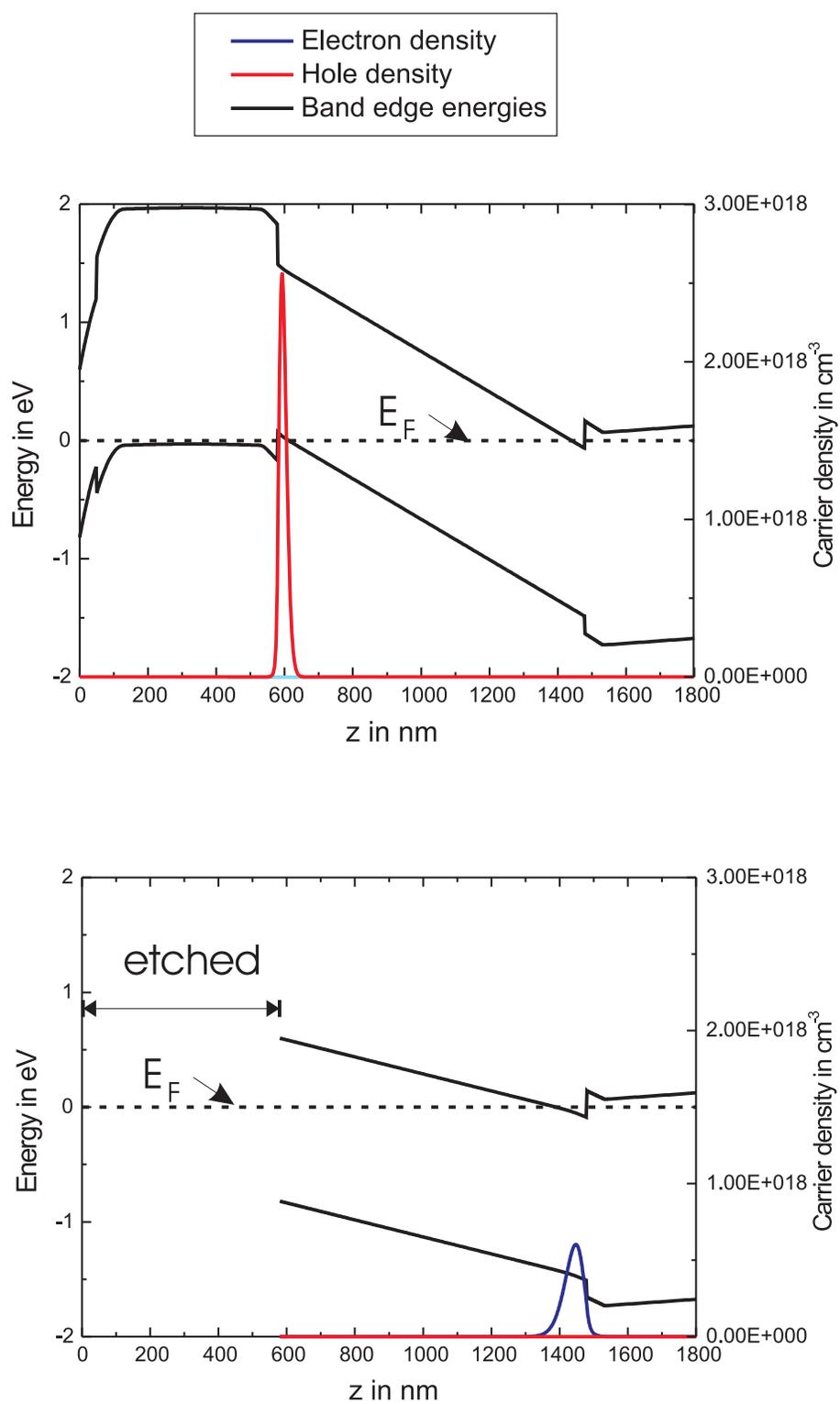}
\end{center}
\caption{\label{figAReal1} Simulated band diagram and charge
density of wafer 1.}
\end{figure}

\begin{figure}[tb]
\begin{center}
\includegraphics[width=12cm]{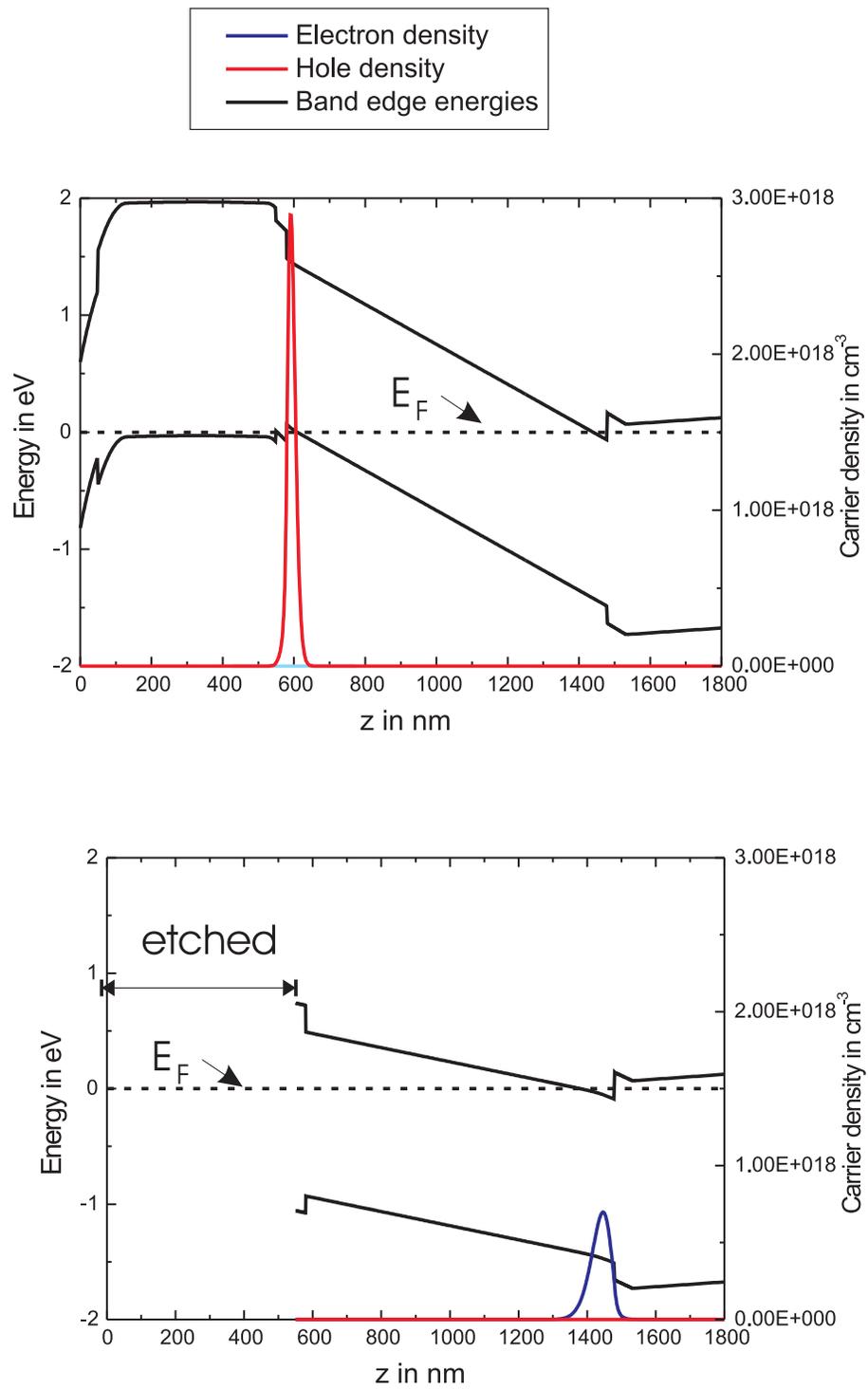}
\end{center}
\caption{\label{figAReal2} Simulated band diagram and charge
density of wafer 2.}
\end{figure}

\clearpage

The donor doping (Silicon) was chosen to be delta-like in order to
allow the implementation of multiple tunnel junction
devices. As a consequence $E_F$ will be pinned at a deep donor level about 70
meV below the conduction band edge, even in the as-grown state. A
narrow GaAs channel would therefore lead to very high electric
fields, close to break down. Hence, the channel width was chosen
to be 90nm, which is three times as large as in Table
\ref{tabAIdeal}. There may be no need for
using a $\delta$-doped Si-layer and the channel can be made
smaller.

A simulation of the band diagrams for both wafers is presented in
Figure \ref{figAReal1} and Figure \ref{figAReal2}. The electron
and hole density is shown as blue and red curves, respectively.
Depletion of the electrons is achieved by the increased
confinement energy in the as-grown state as a result of the
stronger electric field (steeper slope in $E_C(z)$ and $E_V(z)$)
compared to the etched case. The change in charge carrier type,
when the top layers are removed, can be seen in both cases.

The band-diagrams are limiting cases at a sufficiently large
distance along $x$ from the junction. No information on the width
of the junction can be obtained from the 1D Poisson simulation.
The following chapter will investigate the transition region
between these two limiting cases, i.e. the untreated and etched
part of the wafer.
%

\section{Two-Dimensional ATLAS Device Simulation}
\label{secParam2DPoi}

In order to obtain information on the transition between the
band-diagrams in Figure \ref{figAReal1} and Figure
\ref{figAReal2}, the 2D Poisson solver from a device simulation
package (\emph{ATLAS}) from \emph{Silvaco International} was used.
A temperature of 300$\,$K had to be assumed to ensure convergence
of this program. However, the width of the junction should remain
within the same order of magnitude.

\begin{figure}[tb]
\begin{center}
\includegraphics[width=14cm]{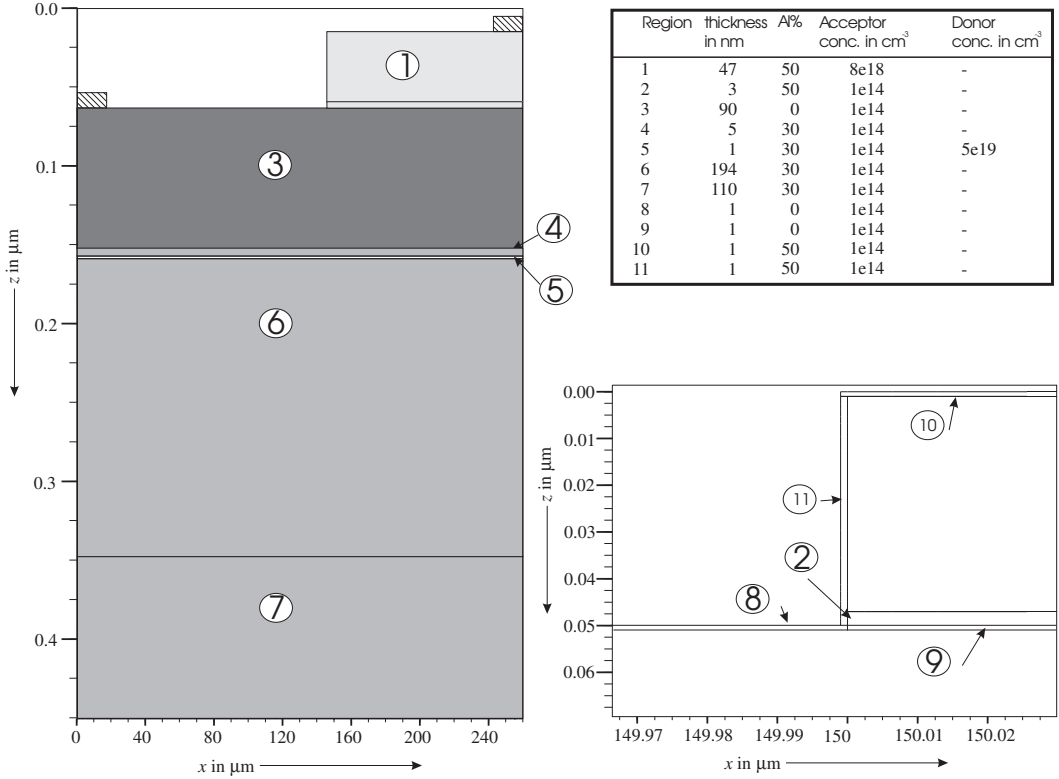}
\end{center}
\caption{\label{figATLASStr} Structure implemented into ATLAS
simulation. The complete shape is shown on the left. Note that
there is a large ratio between the $x$ and $z$ range. Down on the
right the step edge is shown with a larger magnification. The
parameters corresponding to the various layers are given in the
table on the top right.}
\end{figure}

The real structure has to be brought into a format such that
numerical calculations can be performed. First of all, an ideal
sharp corner with vanishing radius is assumed at the step-edge.
Also, in order to have as many mesh points as possible for the
critical regions, the structure has only been simulated to a depth
of about 450nm. Boundary conditions have been set, as described
below to incorporate the effect of layers further down. A complete
layer description can be found in Figure \ref{figATLASStr}. For
region 11 the layer thickness is measured in $x$-direction,
whereas for all other layers it will be in $z$-direction. The
acceptor doping in the real structure will diffuse into the
$i$-AlGaAs (region 2) and $i$-GaAs (region 3) layer. Therefore the
acceptor doping in the simulated structure was allowed to diffuse
such that it drops according to a Gaussian function to half the
original value after 1$\,$nm.

As discussed above, the surface plays a crucial role in the
electrical behavior, because it contains fixed charges that
influence the carriers of the close-by conducting layers. In
\emph{ATLAS} these charges will be calculated directly from the
surface states, instead of fixing $E_F$ as in the 1D simulation.
Here, dangling bonds at the surface can be seen as a sheet of
defects, whose associated energy may lie in the forbidden energy
gap. These so called trap centers exchange charge with the
conduction and valence band through the emission and recombination
of electrons. Two possible states of these traps are assumed here:
empty and full. When empty a trap has a particular cross-section
$\sigma_n$ for capturing an electron. It can either capture or
emit an electron. When the charge on the trap has been changed by
$-q$ by the addition of an electron, it is full, and has a new
cross-section $\sigma_p$ for hole capture. Two basic types of trap
have been found to exist: donor like (electron traps) and
acceptor-like traps (hole traps). The charge contained within each
type of trap will depend upon whether or not an electron or hole
fills the trap.

A donor like trap is positively charged and therefore can only
capture an electron. This means that donor-like traps are {\bf
positive} when {\bf empty} of an electron but are {\bf neutral}
when {\bf filled}. An acceptor trap is negatively charged so
therefore they may only emit an electron. Therefore acceptor-like
traps are {\bf negative} when filled but are {\bf neutral} when
empty.

The probability of occupation assumed by \emph{ATLAS} follows the
analysis by Simmons and Taylor \cite{simmons1PM}. The probability
of occupation is given by the following equations for donor- and
acceptor-like traps, respectively:
\begin{eqnarray}
 F_n & = & \frac{v_n \sigma_n n + e_p}{v_n (\sigma_n n + \sigma_p p) + e_n + e_p},
 \nonumber\\
 F_p & = & \frac{v_p \sigma_p p + e_n}{v_p (\sigma_n n + \sigma_p p) + e_n +
 e_p}\nonumber,
\end{eqnarray}
where $v_n$ and $v_p$ are the thermal velocities for electrons and
holes, and the electron and hole emission rates are given by:
\begin{eqnarray}
 e_n & = & g v_n \sigma_n n_i \exp \frac{E_{DT} - E_i}{k_B T},
 \nonumber\\
 e_p & = & g v_p \sigma_p n_i \exp \frac{E_i - E_{AT}}{k_B T},
\end{eqnarray}
where $k_B$ is the Boltzmann constant, $E_i$ is the intrinsic
Fermi-level position and $g$ is the degeneracy factor of the trap
center. $E_{DT}$ and $E_{AT}$ are the energy level of the donor
trap measured from the valence band edge, and the energy level of
the acceptor trap measured from the conduction band edge,
respectively.

\begin{figure}[tb]
\begin{center}
\includegraphics[width=12.5cm, bb=42 441 534
797]{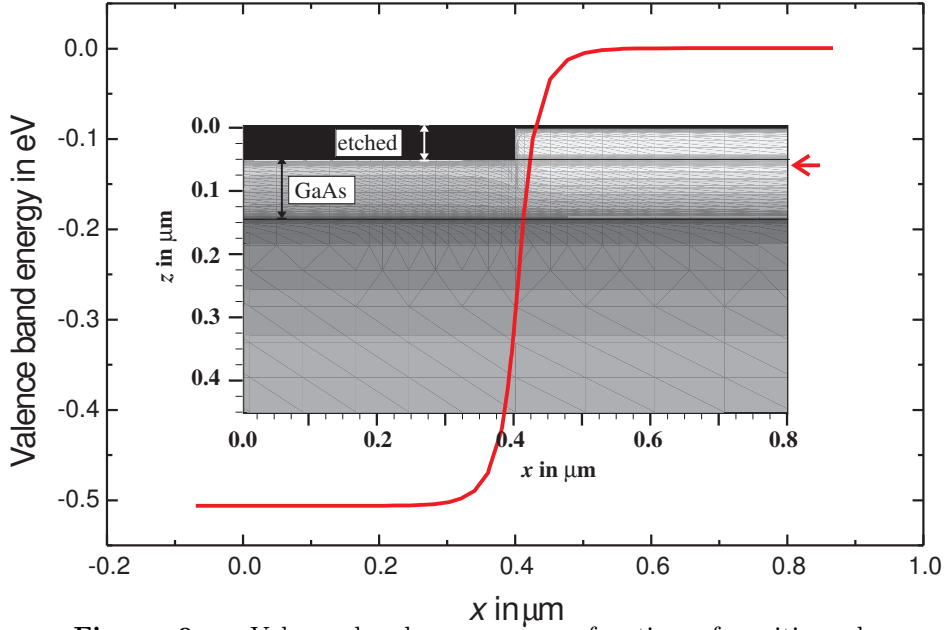}\vspace{-0.9cm}
\end{center}
\caption{\label{fig2DPotSim} Valence band energy as a function of
position along $x$ and $z$ at $T=300$ K simulated by \emph{ATLAS}.
Light values correspond to high energy values. The superimposed
red curve shows the variation of $E_V$ over $x$ for a particular
point $z$, which is indicated by the red arrow.}
\end{figure}

The surface is modelled by assuming a 1$\,$nm thick layer of the
corresponding bulk-material, containing both donor and
acceptor-like traps. The corresponding parameters are:
\begin{center}
\begin{tabular}{|r|r|r|r|r|r|}\hline
  Region & $E_{DT}$ in eV & $E_{AT}$ in eV & $\sigma_n $ in cm$^2$ & $\sigma_p$ in cm$^2$& $\rho_t$ in
  cm$^{-3}$\\\hline \hline
  8 & 0.7 &  -  & 10$^{-8}$ & 10$^{-9}$ & 10$^{-20}$\\
  8 &  -  & 0.7 & 10$^{-9}$ & 10$^{-8}$ & 10$^{-20}$\\
  10 & 0.9 &  -  & 10$^{-8}$ & 10$^{-9}$ & 10$^{-20}$\\
  10 &  -  & 0.9 & 10$^{-9}$ & 10$^{-8}$ & 10$^{-20}$\\
  11 & 0.9 &  -  & 10$^{-8}$ & 10$^{-9}$ & 10$^{-20}$\\
  11 &  -  & 0.9 & 10$^{-9}$ & 10$^{-8}$ & 10$^{-20}$\\ \hline
\end{tabular}
\end{center}
Here $E_{DT}$ measured from the $E_V$ and $E_{AT}$ measured from
$E_C$ are the energy level of the donor and acceptor trap,
respectively.

Apart from the traps the surface regions do not contain any
dopants. Region 11 has been added in order to have the flexibility
to account for different characteristics for different surface
orientations. Region 9 had to be included because otherwise
problems with the automated mesh generation were encountered.
However, this region is exactly the same as the underlying
$i$-GaAs layer. Region 8 and 10 do not contain traps directly
underneath the ohmic contacts. Region 7 has been included  at the
bottom of the structure to pin the Fermi-level at 0.5$\,$eV below
the conduction band edge, the same value as obtained from 1D
simulations.

A two-dimensional plot of the valence band energy $E_V$ is shown
in Figure \ref{fig2DPotSim}. Light areas correspond to high values
of $E_V$. For $x < 0.4\,\mu$m regions 1 and 2 were etched and
therefore the GaAs channel appears darker. The junction width in
$x$ direction was estimated by plotting the variation of $E_V$
over $x$ at the vertical height $z$ indicated by the red arrow. At
this $z$ value the variation in $E_V$ is about $0.5\,$eV, as it
can be seen from the red curve, which is superimposed over the
grayscale plot. The width of the junction is less than 80$\,$nm.
In \cite{kaestner} this value was  verified
experimentally.

\section{Summary}

The concept of carrier type alteration by
selective etching was introduced. This method was then used to
design a lateral junction between an electron and hole gas in a
modulation doped AlGaAs/GaAs heterostructure. The issue of surface
states in the design of this structure was discussed. A
one-dimensional Poisson solver was used to simulate the
banddiagram of various structures at 4.2$\,$K, where carrier type
alteration was predicted when the wafer was etched. The
banddiagram in the transition region was modelled using a
two-dimensional Poisson solver. Because of convergence problems
the 2D simulation had to assume a temperature of 300$\,$K. A
junction width of less than 80$\,$nm along the GaAs channel was
predicted.

\section{References}

\bibliography{PhD}

\begin{thebibliography}{10}

\bibitem{aronov1PAIII}
A.~G. Aronov and G.~E. Pikus.
\newblock Spin injection into semiconductors.
\newblock {\em Sov. Phys. Semicond.}, 10(6):698, 1976.

\bibitem{fiederling1PBII}
R.~Fiederling, M.~Keim, G.~Reuscher, W.~Ossau, G.~Schmidt, A.~Waag, and L.~W.
  Molenkamp.
\newblock Injection and detection of a spin-polarized current in a
  light-emitting diode.
\newblock {\em Nature}, 402:787, December 1999.

\bibitem{ohno1PBII}
Y.~Ohno, D.~K. Young, B.~Beschoten, F.~Matsukura, H.~Ohno, and D.~D. Awschalom.
\newblock Electrical spin injection in a ferromagnetic semiconductor
  heterostructure.
\newblock {\em Nature}, 402:790, 1999.

\bibitem{zhu1PAIII}
H.~J. Zhu, M.~Ramsteiner, H.~Kostial, M.~Wassermeier, H.-P. Sch\"onherr, and
  K.~H. Ploog.
\newblock Room-temperature spin injection from {Fe} into {GaAs}.
\newblock {\em Phys. Rev. Lett.}, 87:016601, 2001.

\bibitem{vaccaro1PBII}
P.~O. Vaccaro, H.~Ohnishi, and K.~Fujita.
\newblock A light-emitting device using a lateral junction grown by
  molecular-beam epitaxy on {GaAs (311)} a-oriented substrates.
\newblock {\em Appl. Phys. Lett.}, 72(7):818, 1998.

\bibitem{kaestner1PBII}
B.~Kaestner, D.~G. Hasko, and D.~A. Williams.
\newblock Lateral p-n junction in modulation doped {AlGaAs/GaAs}.
\newblock {\em Jpn. J. Appl. Phys.}, 41:2513--2515, 2002.

\bibitem{cecchini1PBII}
Marco Cecchini, Vincenzo Piazza, Fabio Beltram, Marco Lazzarino, M.~B. Ward,
  A.~J. Shields, H.~E. Beere, and D.~A. Ritchie.
\newblock High-performance planar light-emitting diodes.
\newblock {\em Appl. Phys. Lett.}, 82(4):636, 2003.

\bibitem{miller2PBII}
D.~L. Miller.
\newblock Lateral $p-n$ junction formation in gaas molecular beam epitaxy by
  crystal plane dependent doping.
\newblock {\em Appl. Phys. Lett.}, 47(12):1309, 1985.

\bibitem{yamamoto1PBII}
T.~Yamamoto, M.~Inai, M.~Hosoda, T.~Takabe, and T.~Watanabe.
\newblock Demonstration of lateral p-n subband junctions in {Si}-delta-doped
  quantum-wells on {(111)A} patterned substrates.
\newblock {\em Jpn. J. Appl. Phys.}, 32(10):4454, 1993.

\bibitem{bastard1PBII}
Gerald Bastard.
\newblock {\em Wave mechanics applied to semiconductor heterostructures}.
\newblock Monographies de Physique. Halsted Press, New York, 1988.

\bibitem{ashcroft1PBII}
Neil~W. Ashcroft and N.~David Mermin.
\newblock {\em Solid State Physics}.
\newblock Saunders College Publishing, 1976.

\bibitem{cardona1PBII}
P.~Y. Yu and M.~Cardona.
\newblock {\em Fundamentals of Semiconductors}.
\newblock Springer, Berlin, New York, 3rd edition, 2001.

\bibitem{spicer1PM}
W.~E. Spicer, P.~W. Chye, P.~R. Skeath, C.~Y. Su, and I.~Lindau.
\newblock New and united model for {Schottky} barrier and {III-V} insulator
  interface state formation.
\newblock {\em J. Vac. Sci. Technol. B}, 16:1422, 1979.

\bibitem{thurmond1PM}
C.~D. Thurmond, G.~P. Schwartz, G.~W. Kammlotot, and B.~Schwartz.
\newblock {\em J. Electrochem. Soc.}, 127:1366, 1980.

\bibitem{ping1PM}
J.~G. Ping and H.~E. Ruda.
\newblock The origin of {Ga$_2$O$_3$} passivation for reconstructed {GaAs(001)}
  surfaces.
\newblock {\em J. Appl. Phys.}, 83:5880, 1998.

\bibitem{yi1PM}
S.~I. Yi, P.~Kruse, M.~Hale, and A.~C. Kummela.
\newblock Adsorption of atomic oxygen on {GaAs(001)}-(2$\times$4) and the
  resulting surface structures.
\newblock {\em J. Chem. Phys.}, 114(7):3215, 2001.

\bibitem{hasegawa1PM}
H.~Hasegawa and T.~Sawada.
\newblock {\em Thin Solid Films}, 103:119, 1983.

\bibitem{tan1POI}
I.-H. Tan, G.~L. Snider, L.~D. Chang, and E.~L. Hu.
\newblock A self-consistent solution of {Schr\"odinger-Poisson} equations using
  a nonuniform mesh.
\newblock {\em J. Appl. Phys.}, 68(8):4071, 1990.

\bibitem{wu1POI}
X.~S. Wu, L.~A. Coldren, and J.~L. Merz.
\newblock Selective etching characteristics of {HF} for
  $\mathrm{Al_xGa_{1-x}As/GaAs}$.
\newblock {\em Electronics Letters}, 21(13):558, 1985.

\bibitem{kaestner2PBII}
B.~Kaestner, D.~G. Hasko, and D.~A. Williams.
\newblock Quasi-lateral {2DEG-2DHG} junction in {AlGaAs/GaAs}.
\newblock {\em Microelectronics Journal}, 34(5-8):423, 2003.

\bibitem{kaestner3}
B.~Kaestner, D.~A. Williams, and D.~G. Hasko.
\newblock Nanoscale lateral light emitting p-n junctions in algaas/gaas.
\newblock {\em Microelectronics Engineering}, 67-8:797, 2003.

\bibitem{kaestner}
B.~Kaestner, C.~Sch\"onjahn, and C.~J. Humphreys.
\newblock Mapping the potential within a nanoscale undoped gaas region using a
  scanning electron microscope.
\newblock {\em Appl.Phys.Lett.}, 84(12):2109, 2004.

\bibitem{parikh1POI}
P.~A. Parikh, S.~S. Shi, J.~Ibettson, E.~L. Hu, and U.~K. Mishra.
\newblock Hydrogeneration of {GaAs MISFETs} with {Al$_2$O$_3$} as the gate
  insulator.
\newblock {\em Electr. Lett.}, 32(18):1724, 1996.

\bibitem{simmons1PM}
J.~G. Simmons and G.~W. Taylor.
\newblock Nonequilibrium steady-state statistics and associated effects for
  insulators and semiconductors containing an arbitrary distribution of traps.
\newblock {\em Phys. Rev. B}, 4:502, 1971.

\end{thebibliography}

\end{document}